\documentclass[aps,prd,twocolumn,showpacs]{revtex4-1}

\usepackage{float,graphicx,caption}

\usepackage[hidelinks]{hyperref}

\begin{document}

\title{Dark sterile neutrinos on a linear seesaw of neutrino masses}
\author{\footnotesize Ernesto A. Matute}
\email{ernesto.matute@usach.cl}
\affiliation{Departamento de F\'{\i}sica, Facultad de Ciencia,
Universidad de Santiago de Chile (USACH), Chile}

\begin{abstract}
Sterile neutrinos as source of mass and flavor mixing of active neutrinos as well as genesis
of dark matter (DM) and matter-antimatter asymmetry have gained special interest. Here we study
the case of the Standard Model (SM) extended with three right-handed (RH) neutrinos and a dark
sector with two extra sterile neutrinos, odd under a discrete $Z_2$ symmetry. The RH neutrinos
are responsible for producing baryon asymmetry via high-scale unflavored leptogenesis. They are
superheavy and their abundance at the electroweak broken stage is vanishingly small, so that
they have no impact on phenomenology at low energies. The two dark neutrinos generate the tiny
mass of two active neutrinos through a mechanism similar to the minimal linear seesaw, and
saturate the relic abundance as freeze-in DM coming mainly from decays of SM weak-gauge bosons
via active-dark neutrino mixing. The absence of the dark Majorana mass terms in the dark linear
seesaw is explained by invoking a hidden symmetry, the so-called presymmetry, and the DM
candidate appears in the form of a quasi-Dirac neutrino. The $Z_2$ symmetry is broken in the
dark neutrino sector, but exact in the realm of RH neutrinos. The required coupling weakness
for the freeze-in DM neutrino is related to a very small breach of the unitarity of the active
neutrino mixing matrix. We show how phenomenological constraints on the production and decay of
the DM neutrino imply an upper bound around 1 MeV for its mass and unitarity up to
$\mathcal{O}(10^{-7})$ for the mixing matrix.

\vspace{8pt}
\noindent \emph{Keywords}: Neutrino mass; dark sterile neutrino; dark linear seesaw; superheavy
RH neutrino; presymmetry.
\end{abstract}

\maketitle

\section{Introduction}
\label{Introduction}

Neutrino mass and flavor mixing \cite{N1,N2}, as well as the existence of dark matter
(DM) \cite{DM} and baryon asymmetry of the universe \cite{DMrelic}, are overwhelming evidences
for physics beyond the Standard Model (SM). The finding of an extension of the SM where these
phenomena appear connected and explained in terms of a relatively few number of new massive
particles is a captivating possibility.

The most economic choice is that of just three right-handed (RH) neutrinos as in the type-I
seesaw scenario \cite{Shapo1,Shapo2}. This option is constrained to a region having mass in
the keV-MeV range for the lightest one, identified as a freeze-in type of DM, and mass scale
near $10^{10}$ GeV for the two other RH neutrinos to address the origin of the tiny mass of
the active left-handed (LH) neutrinos of the SM and the baryon asymmetry of the universe via
flavored leptogenesis \cite{segregative}, segregating and exacerbating differences among the
three RH neutrinos. It may be ratified or ruled out by upcoming experiments
\cite{whitepaper,sterileNe}, adding motivation to look for alternative scenarios with extra
particles as well as symmetry enlargements.

In this paper we consider a new variant that differs in some respect from the type-I seesaw
mechanism referred to above. It also deals with a beyond the SM context that aims at describing
active neutrino mass and mixing, the presence of a DM candidate and baryogenesis through
leptogenesis. The SM spectrum is enlarged with three RH neutrinos and two exotic leptons that
we call sterile neutrinos; all of them are singlets under the SM gauge symmetries.
Regarding extra symmetries, we invoke the so-called presymmetry \cite{Presym} and a related
discrete $Z_2$ symmetry that distinguishes the extra sterile neutrinos from the presymmetric
fermions. More specifically, presymmetry is a hidden lepton-quark symmetry in the frame of
the SM extended with RH neutrinos. It is an underlying symmetry based on the requirement
of left-right symmetry of fermionic content and symmetric fractional electroweak charges,
postulated ab initio with a global $U(1)_{B-L}$ symmetry which forbids Majorana mass terms.
Presymmetry is broken at the level of standard leptons and quarks. Yet, it provides a natural
linking of the active LH neutrinos with their RH partners prior to the breaking that split
them up. It is the origin of the $Z_2$ symmetry for the new leveling of singlet fermions.

The dark sector we propose contains the two mentioned extra sterile neutrinos, which are odd
under the discrete $Z_2$ symmetry, whereas all SM particles and the RH neutrinos are even under
the same symmetry. These new sterile neutrinos, which provide the DM of the universe, break
the $Z_2$ symmetry to generate the active neutrino masses via a dark linear seesaw mechanism,
the lightest of them being massless at tree level. We make use of presymmetry to understand
the absence of their Majorana mass terms, in contrast to supersymmetry as originally adduced
in the context of the usual linear seesaw model \cite{Linear1,Linear2,Linear3}, so
avoiding to rely on a plethora of unobserved particles and interactions. It may be said that
presymmetry helps to configure the low-energy physics, while supersymmetry does it at the
very high scale if it exists.

Since current experiments do not show evidence of effects from extensions of the SM with RH
neutrinos, we assume that all of them are too heavy and their abundance at the electroweak
broken stage extremely small to have significant influence on the low-energy phenomenology,
meaning a parameter space not bounded by the low-energy physics, with the result that they
are neither responsible for neutrino mass generation nor constituent part of DM. We adhere,
however, to the paradigm that the three heavy RH neutrinos are the source of the baryon
asymmetry of the universe via high-scale leptogenesis \cite{leptogenesis}, a mechanism where the
three RH neutrinos decay into the lepton doublet and the Higgs doublet by means of the Yukawa
interactions. These processes, assumed out-of-thermal-equilibrium in the early universe, are
lepton-number-violating as the RH neutrinos are their own antiparticles and CP-violating due to
the asymmetry between such a decay and its CP-conjugate process involving the antiparticles.
Since the extra sterile neutrinos take care of the light neutrino masses and therefore no mass
hierarchy within the super-heavy RH neutrinos is required, our best choice is the unflavored
leptogenesis with no connection between low and high energy CP
violations \cite{NoCP1,NoCP2,NoCP3}. The RH neutrino masses are so large that all
the relevant Yukawa interactions leading to the final lepton asymmetry are unable to distinguish
among lepton flavors \cite{Leptogenesis1,Leptogenesis2,Leptogenesis3}. In our setting, the
leptogenesis and the neutrino mass generation work at different scales, in contrast to the usual
construction where they have a same scale without involving the extra sterile neutrinos and
assuming a flavored leptogenesis \cite{segregative}. On the other hand, the relic abundance of
heavy RH neutrinos, necessary to have an acceptable lepton asymmetry, introduces constraints
on scenarios proposed for the very early universe
\cite{AfterInfla1,AfterInfla2,AfterInfla3,AfterInfla4,AfterInfla5}.

Quantitatively, we can apply the usual type-I seesaw formula
$m_\nu = m^2_D / M_R$ with $M_R \sim 10^{13}$ GeV for RH neutrino masses to see the viability
of unflavored leptogenesis in our scenario \cite{ZZX}. From the point of view of model building,
a natural expectation is a Dirac neutrino mass $m_D$ similar in size to the Dirac mass of a
charged lepton, i.e., $m_D \sim m_\tau \sim \mbox{1 GeV}$ for the third generation of neutrinos,
which leads to $\Delta m_\nu \sim 10^{-5}$ eV. In this context, the size of their Yukawa
couplings is given by $y_\nu = \sqrt{2} m_D / v_\phi \sim 10^{-3}$ for $v_\phi = \mbox{246 GeV}$.
These results change to $\Delta m_\nu \sim 10^{-7}$ eV and $y_\nu \sim 10^{-4}$
if the similarity condition involving the second generation of charged leptons is considered
instead, i.e., $m_D \sim m_\mu \sim 10^2$ MeV. This simple quantitative argument implies
that we can have desired unflavored leptogenesis with tiny Yukawa couplings while keeping
the corresponding contribution to the mass of the two massive active neutrino suppressed.
In the case of the first generation, $\Delta m_\nu \sim 10^{-12}$ eV and $y_\nu \sim 10^{-6}$
for $m_D \sim m_e \sim 1$ MeV.

Even though RH neutrinos decouple from the low-scale phenomenology and their introduction and role
appear to be superfluous with respect to the main topic of the paper, focused on the neutrino mass
and DM issues, they are necessary to have presymmetry and so the dark linear seesaw mechanism.
Thus, an interplay between active neutrinos, dark neutrinos, and RH neutrinos can explain the
tiny active neutrino masses, the observed DM relic abundance, and the matter-antimatter asymmetry
of the universe. No new gauge interactions beyond the SM are considered. It is a compelling way
to relate these phenomena, where the dark sterile neutrinos of the low-scale seesaw mechanism
provide a DM candidate and the RH neutrinos of the high-scale leptogenesis have practically no
effects on the low-energy phenomenology because of their huge mass and vanishingly small abundance
at the electroweak broken phase.

Concerning DM in the early universe when temperature was extremely high, the scenario
contemplates a thermal bath of the SM particles with the RH neutrinos propagating and
decaying (into lepton and Higgs doublets) out of equilibrium as they carry no gauge
charge of the SM, while DM is assumed to be produced non-thermally via decay of the SM
weak-gauge and Higgs bosons with a zero or negligible initial abundance and very feeble
coupling to the thermal bath, the freeze-in mechanism \cite{freezein1,freezein2} with the
RH neutrino marginalizing the dark neutrino because of its charge under the $Z_2$ symmetry.
The very small mixing between active and dark neutrinos required to achieve this picture
will be related to the tiny breach of the unitarity of the active neutrino mixing matrix
and connected to a minuscule mass parameter that in the Neutrino Minimal Standard Model
($\nu$MSM) \cite{Shapo1,segregative}, with three RH neutrinos only, is interpreted as
the mass of the lightest active neutrino. Surprisingly large thermal one-loop corrections
that suppress active-dark neutrino mixings at the time of dark neutrino production,
computed with the thermal quantum field theory, are not taken into account as they are
not perturbative at all \cite{thermalQFT}.

Here we revisit the model presented in~\cite{EAM1} to consider the DM sterile neutrino
above the low keV scale, without involving the Dodelson-Widrow production mechanism \cite{DW} and
its Shi-Fuller variant \cite{SF}. With the RH neutrinos decoupled from the low-scale phenomenology
due to their heaviness and vanishing abundance, the two dark sterile neutrinos become essential to
provide active neutrino masses via the seesaw mechanism, and also indispensable to fulfill the
observed DM relic abundance in a minimal way. Given this new scenario, we claim that the active
neutrinos get their tiny mass from the dark sector through a low-scale dark linear seesaw, breaking
the dark $Z_2$ symmetry, where none of the three RH neutrinos are engaged in. Current constraints
place the DM neutrino mass in the keV to MeV range, motivating thus the search for sterile neutrinos
at the sub-MeV scale.

Sub-MeV massive sterile neutrinos, on the other hand, have been proposed in the literature under
various scenarios, as done for example in Refs.~\cite{Shapo1,segregative,Model1,Model2,Model3,Model4,Model5,Model6}.
However, they are different from ours from the point of view of both the model and the phenomenology.
In those proposals either RH Majorana neutrinos are considered or extra sterile neutrinos, with no
involvement in neutrino mass generation, are included. Our approach would be ruled out if,
for instance, active neutrinos turn out to be Dirac or pseudo-Dirac fermions, or RH weak
currents are discovered at low energies, validating the left-right symmetric gauge theories
\cite{LRsym1,LRsym2,LRsym3}.

In a sense, our two extra sterile neutrinos are RH neutrinos that generate neutrino masses with
so much small couplings that appear to contravene previous calculations. However, such a space
parameter is allowed because of the much lower seesaw scale, which enables to have a DM
candidate from the extra sterile neutrinos. It contrasts with known models that relate neutrino
mass generation and DM using different particles and leading to a profusion of phenomenological
signatures, as in Refs.~\cite{otherDM1,otherDM2,otherDM3,otherDM4}.

The paper is organized as follows. In Sec.~\ref{Model}, we revisit the model, describing the
realization of the dark linear seesaw with just two extra sterile neutrinos through which the
active neutrinos get their tiny mass, and having a better understanding of the weakness of the
active-dark neutrino mixing necessary for the freeze-in process. In Sec.~\ref{Phenomenology},
we discuss on the production and decay mechanisms of the DM neutrino as well as the
phenomenological constraints from DM searches at low energies. Potentially large thermal
one-loop corrections that suppress active-dark neutrino mixings, calculated with the thermal
quantum field theory at temperatures around the electroweak scale, are not included as they do
not have the expected perturbative values. The results presented here show that such sterile
neutrinos with such small couplings can be responsible for the active neutrino masses and at
the same time provide a valid DM candidate. We conclude in Sec.~\ref{Conclusion}.

\section{The model}
\label{Model}

The SM is extended with three RH neutrinos ($\nu_R$) and a dark sector that contains
two extra sterile neutrinos that we denote as ($N_{1R},N_{2L}$). These dark fields,
singlets of the SM gauge group, are odd under the discrete $Z_2$ symmetry, while the SM
particles and the RH neutrinos are even. The RH neutrinos restore left-right symmetry in the
neutrino content of the SM, but not an interchange left-right symmetry as usually imposed
in left-right symmetric gauge models. They deal with baryon asymmetry via the high-scale
unflavored leptogenesis, whereas the dark sterile neutrinos generate light neutrino masses
through a low-scale variant of the usual linear seesaw mechanism, breaking the $Z_2$
symmetry, and provide a freeze-in DM neutrino candidate via active-dark neutrino mixing.
This dark neutrino model \cite{EAM1} is revisited in the following.

The RH neutrinos are so heavy because of their Majorana mass terms, and their abundance at
the electroweak broken stage so small as result of their decays at high energies, that they
have no impact in the low-energy physics and, in particular, any mixing with active and dark
neutrinos is effectively reduced to nothing. As noted in Sec.~\ref{Introduction}, the RH
neutrino contribution to active neutrino mass generation via type-I seesaw is strongly
suppressed in the scenario of an unflavored leptogenesis. For this reason, we focus our
attention to the effects of the extra sterile neutrinos, without considering RH neutrinos.
We start with the effective seesaw Lagrangian~\cite{EAM1}
\begin{equation}
- \mathcal{L} \supset {\it y}^\prime_\nu \overline{\ell_L} \,
\widetilde{\phi} N_{1R} + {\it y}^\prime_L \overline{\ell_L} \,
\widetilde{\phi} N^c_{2L} + M_D \overline{N_{2L}} N_{1R} + h.c. ,
\label{neutrinos}
\end{equation}
where the $Z_2$ symmetry is broken in terms containing just one dark field.
After the electroweak symmetry breaking, the neutrino 5$\times$5 mass matrix
in the basis ($\nu_L$, $N^c_{1R}$, $N_{2L}$) of active and dark neutrinos is as follows,
\begin{equation}
\mathcal{M}_\nu = \left(
\begin{array}{ccc}
0 & m^\prime_D & \mu^\prime_L \\ && \\ m^{\prime T}_D & 0 & M_D \\ && \\
\mu^{\prime T}_L & M_D & 0
\end{array} \right) ,
\label{massmatrix}
\end{equation}
where
\begin{equation}
m^\prime_D = \frac{\it{y}^\prime_\nu v_{\phi}}{\sqrt{2}} , \quad
\mu^\prime_L = \frac{\it{y}^\prime_L v_{\phi}}{\sqrt{2}} ,
\label{origmass}
\end{equation}
with $v_{\phi}=$ 246 GeV, are 3$\times$1 submatrices, while $M_D$ is a 1$\times$1
submatrix. The mass hierarchy is $m^\prime_D , \mu^\prime_L \ll M_D$,
with $m^\prime_D \sim \mu^\prime_L$, attributable to the breaking of $Z_2$ symmetry.
The second and third diagonal elements of the mass matrix $\mathcal{M}_\nu$ in
Eq.~(\ref{massmatrix}) come to be zero because of the underlying presymmetry discussed
in Sec.~\ref{Introduction}. Presymmetry is broken in couplings involving the LH and RH
neutrinos, allowing heavy Majorana mass terms for the latter and small $m^\prime_D$, $\mu^\prime_L$
breaking terms for the former. However, Majorana mass terms for the extra sterile neutrinos are
still forbidden.

The mass matrix can be diagonalized by a unitary matrix $U$, such that
\begin{equation}
U^T \mathcal{M}_\nu U = \mbox{diag} (0, m_2, m_3, M_1, M_2) ,
\label{diag}
\end{equation}
where $m_1=0$, $m_2$, $m_3$ are the three lighter eigenvalues of $\mathcal{M}_\nu$
in the normal ordering (or $m_3=0$, $m_1$, $m_2$ in the inverted ordering, changing
subindexes 1, 2, 3 by 3, 1, 2, respectively, in Eq.~(\ref{diag}) and those that follow
below), and $M_1, M_2$ are the heavier ones. They are approximately given by the seesaw
forms
\begin{eqnarray}
\displaystyle m_\nu &=& - \frac{m^\prime_D \mu^{\prime T}_L}{M_D} -
\frac{\mu^\prime_L m^{\prime T}_D}{M_D} , \nonumber \\ && \nonumber \\
\displaystyle M_1 &=& M_D + \frac{(m^\prime_D + \mu^\prime_L)^T
(m^\prime_D + \mu^\prime_L)}{2 M_D} , \nonumber \\ && \nonumber \\
\displaystyle M_2 &=& - M_D - \frac{(m^\prime_D - \mu^\prime_L)^T
(m^\prime_D - \mu^\prime_L)}{2 M_D} ,
\label{eigenvalues}
\end{eqnarray}
obtained from the block diagonalization of $\mathcal{M}_\nu$. The neutrino mass
$m_\nu$ is a 3$\times$3 matrix, while $M_1$ and $M_2$ are numbers. Within this
scenario, one of the active neutrinos is massless at tree level, while the two
others obtain their tiny masses from the dark sector, as shown in Ref.~\cite{EAM2}.
Entries for $m^\prime_D$ and $\mu^\prime_L$ leading to the mass eigenvalues
$(0, m_2, m_3)$ are given below. Note that the mass splitting between the two dark
neutrinos depends on the mass of active neutrinos; the neutrino masses prevent the
degeneracy of the two sterile neutrinos what otherwise would appear as a Dirac
fermion. Also, as expected from the null trace of
$\mathcal{M}_\nu$, $\mbox{tr}(m_\nu)+M_1+M_2=0$.

It is worth mentioning that the mass matrix in Eq.~(\ref{massmatrix}) and
its eigenvalues in Eq.~(\ref{eigenvalues}) have structures similar to the ones
discussed in the standard linear seesaw, but with $m^\prime_D$ instead of
$m_D$, where $m_D = \it{y}_\nu v_\phi / \sqrt{\mbox{2}}$ with $\it{y}_\nu$
being the regular Yukawa couplings involving the RH neutrinos. This new
framework implies a much lower scale. Specifically, with the mass hierarchy
$m^\prime_D \sim \mu^\prime_L \ll m_D$, there exists an extra suppression
with respect to the usual linear seesaw given essentially by
\begin{equation}
m_\nu =  \left( \frac{2 m_D \mu_L^\prime}{M_D} \right) \frac{m_D^\prime}{m_D} ,
\end{equation}
where $m_D^\prime / m_D$ is the suppressing factor leading to the form in
Eq.~(\ref{eigenvalues}). The scale of $M_D$ can then be brought down even
to the keV scale for $\mu_L^\prime$ and $m_D^\prime$ at the eV range, associated
with the breaking of the $Z_2$ symmetry and independently of any similarity
condition that can be established between the size of the Dirac neutrino mass
$m_D$ and the Dirac mass of a charged lepton. Thus, the parameter space obtained
in previous studies in terms of $m_D$ (involving RH neutrinos) is not the
same to that computed in our context with $m_D^\prime$ (involving dark
neutrinos). Moreover, the absence of the 22 and 33 terms in Eq.~(\ref{massmatrix})
is explained by invoking the hidden presymmetry, while in the standard linear
seesaw setting has been done by adducing supersymmetry and so a large amount
of new particles and interactions, none of which has been observed.
Alternatively, there is also the possibility of such a removal in
non-supersymmetric versions of the regular linear seesaw. But, a more
involved particle content is required \cite{nonSUSY1,nonSUSY2}.

The mass eigenstates ($\nu_i,N_1,N_2$), $i=1,2,3$, are related to the active and dark
neutrinos ($\nu_\alpha,N_{1R}^c,N_{2L}$), $\alpha=e,\mu,\tau$, by the unitary matrix $U$ as
\begin{equation}
\left( \begin{array}{c}
\nu_\alpha \\ \\ N_{1R}^c \\ \\ N_{2L}
\end{array} \right) = U
\left( \begin{array}{c}
\nu_i \\ \\ N_1 \\ \\ N_2
\end{array} \right) .
\label{totalmixing}
\end{equation}

Neutrino oscillation data provide values  very close to those expected from a unitary
matrix linking the active neutrino states \cite{PDG}. Thus, as a starting point, we
follow the unitary approach to the active neutrino mixing matrix. This means to neglect
mixing between the mass eigenstates of active neutrinos ($\nu_i$) and dark neutrinos
($N_1, N_2$) to begin with, even though a tiny coupling is necessary to generate the
neutrino masses.

Now, since CP violating phases are currently unknown \cite{CPphase}, and that the CP-conserving
regime is still admitted, here we take the case of a real mixing matrix. It is then shown
from Eq.~(\ref{eigenvalues}) that the entries for the elements of the neutrino mass matrix $m_\nu$
satisfy the condition
\begin{equation}
\sum_{\alpha=e,\mu,\tau} (|m^\prime_{D\alpha}|^2+|\mu^\prime_{L\alpha}|^2) = M_D (m_2+m_3) .
\label{sumrule}
\end{equation}
For instance, assuming symmetry between $\mu$ and $\tau$ flavors of the second and
third generation of neutrinos, we have the following elements in the first place:
\begin{eqnarray}
\displaystyle \frac{m^\prime_{De}}{\sqrt{M_D}} &=& \sqrt{\frac{m_2}{6}} =
\frac{\mu^\prime_{Le}}{\sqrt{M_D}} , \nonumber \\ && \nonumber \\
\displaystyle \frac{m^\prime_{D\mu}}{\sqrt{M_D}} &=& \sqrt{\frac{m_2}{6}} +
i \frac{\sqrt{m_3}}{2} = \frac{\mu^{\prime *}_{L\mu}}{\sqrt{M_D}} , \nonumber \\ && \nonumber \\
\displaystyle \frac{m^\prime_{D\tau}}{\sqrt{M_D}} &=& \sqrt{\frac{m_2}{6}} -
i \frac{\sqrt{m_3}}{2} = \frac{\mu^{\prime *}_{L\tau}}{\sqrt{M_D}} ,
\label{TBMmixings}
\end{eqnarray}
with $|m^\prime_{D\alpha}|=|\mu^\prime_{L\alpha}|$, which is a characteristic of
our dark linear seesaw. In this way, the massless of the lightest active neutrino,
the nonzero mass values of the two others, and an appropriate flavor mixing matrix
are guaranteed. Details for these neutrino couplings are found in Ref.~\cite{EAM2},
where the tri-bimaximal (TBM) mixing \cite{TBM} and its deviations are derived
straightforwardly on the basis of the symmetry between $\mu$ and $\tau$ flavors.

Yet, the unitarity of the active mixing matrix $U_{\alpha i}$ is expected to be
violated by means of tiny, effective active-dark neutrino mixing terms, as
\begin{equation}
\nu_\alpha = U_{\alpha i} \nu_i + U_{\alpha N_1} N_1 + U_{\alpha N_2} N_2 ,
\end{equation}
where
\begin{equation}
U_{\alpha N_1} = \frac{\Delta m_{D \alpha}^{\prime}}{M_D} , \quad
U_{\alpha N_2} = \frac{\Delta \mu_{L \alpha}^{\prime}}{M_D} ,
\label{darkmixing}
\end{equation}
with $\Delta m_{D \alpha}^{\prime}$ and $\Delta \mu_{L \alpha}^{\prime}$ being
small effective mass terms to represent changes in the values of
$m_{D \alpha}^{\prime}$ and $\mu_{L \alpha}^{\prime}$ of the original seesaw
Lagrangian due to nonunitarity of active neutrino mixing matrix and
active-dark neutrino mixing, as shown in the following. Mixings in
Eq.~(\ref{darkmixing}) contrast with the rough values of order
$m_{D \alpha}^{\prime}/M_D$ and $\mu_{L \alpha}^{\prime}/M_D$ obtained from the
block diagonalization of the mass matrix $\mathcal{M}_\nu$ in Eq.~(\ref{massmatrix}),
with an approximately unitary mixing matrix as a first step. As we shall see, they
are independent from neutrino masses in Eq.~(\ref{TBMmixings}) by which the
active neutrino mass matrix in Eq.~(\ref{eigenvalues}) becomes diagonalizable with
the unitary TBM matrix, and accordingly they are several orders of magnitude below
those ones, given the constraints on DM abundance.

Using the unitary argument, we extent the pattern of Eq.~(\ref{sumrule})
to include the tiny terms of Eq.~(\ref{darkmixing}):
\begin{eqnarray}
&&\sum_\alpha (|m^\prime_{D\alpha}|^2)+|\mu^\prime_{L\alpha}|^2+
M_D^2 |\theta|^2 = \sum_\alpha (|m^\prime_{D\alpha}|^2+|\mu^\prime_{L\alpha}|^2
\nonumber \\
&& \hspace{0.5cm} + |\Delta m^{\prime}_{D\alpha}|^2+|\Delta \mu^{\prime}_{L\alpha}|^2)
= M_D (m_2+m_3 + m),
\label{sumruleext}
\end{eqnarray}
having the effective active-dark neutrino mixing
\begin{equation}
|\theta|^2= \sum_\alpha \frac{|\Delta m^{\prime}_{D\alpha}|^2+|\Delta \mu^{\prime}_{L\alpha}|^2}{M_D^2}
= \frac{m}{M_D} .
\label{theta}
\end{equation}
This quantity, which in our context parameterizes the departure from unitarity of the
active neutrino mixing matrix and the active-dark neutrino mixing, characterizes the
phenomenology of the DM neutrino. For the sake of simplification, we opt here for
symmetric values involving all flavors:
\begin{equation}
\frac{\Delta m^{\prime}_{D\alpha}}{\sqrt{M_D}} =
\frac{\Delta \mu^{\prime}_{L\alpha}}{\sqrt{M_D}} = \sqrt{\frac{m}{6}} ,
\label{effval}
\end{equation}
so that their contributions in Eq.~(\ref{sumruleext}) are indeed extremely limited,
compared to those of $m^{\prime}_{D\alpha}$ and $\mu^{\prime}_{L\alpha}$ in
Eq.~(\ref{TBMmixings}).

Thus, from Eq.~(\ref{effval}),
\begin{equation}
\displaystyle \Delta m_{D \alpha}^{\prime} = \Delta \mu_{L \alpha}^{\prime} =
\sqrt{\frac{m M_D}{6}} ,
\label{dev}
\end{equation}
and, going back to Eq.~(\ref{darkmixing}),
\begin{equation}
U_{\alpha N_1} = U_{\alpha N_2} = \sqrt{\frac{m}{6 M_D}} .
\label{findarkmix}
\end{equation}
As expected, $\Delta m_{D \alpha}^{\prime} , \Delta \mu_{L \alpha}^{\prime} \ll
m_{D \alpha}^{\prime} , \mu_{L \alpha}^{\prime}$, even though the rough approximation
of Eq.~(\ref{effval}) is not consistent with unitarity of the whole 5$\times$5 mixing
matrix. Consequently, the tiny mass $m$ can be considered as vestige of very small
effective active-dark neutrino mixings; in Refs.~\cite{Shapo1,segregative}, it is
interpreted as the mass of the lightest active neutrino (i.e., $m=m_1$).

All in all, the magnitude $|m^{\prime}_{D\alpha}|=|\mu^{\prime}_{L\alpha}|$ is related
to the neutrino masses $m_2$ and $m_3$, while that of
$|\Delta m^{\prime}_{D\alpha}|=|\Delta \mu^{\prime}_{L\alpha}|$ is connected with the
tiny mass $m$, assuming the normal ordering. In the context of the dark seesaw with only
two dark neutrinos, however, $m$ cannot be associated with the lightest active neutrino,
so that a value at a very, very small scale is expected, perturbing the nonzero neutrino
masses. In our framework, it is attributed to the nonunitarity of the standard neutrino
mixing matrix and the active-dark neutrino mixing. We have checked that the 3$\times$3
block matrix of active neutrino masses, after including in Eqs.~(\ref{massmatrix}) and
(\ref{eigenvalues}) terms in Eq.~(\ref{TBMmixings}) and contributions from dark neutrinos
according to the first approximation of Eq.~(\ref{dev}), is diagonalized by the unitary
TBM matrix with eigenvalues $(0, m_2+2\sqrt{m_2 m}, m_3)$ up to leading order in $m$.
The addition patterns of Eqs.~(\ref{sumrule}) and (\ref{sumruleext}) are generalized to
\begin{eqnarray}
&& \sum_\alpha (|m^\prime_{D\alpha}+\Delta m^{\prime}_{D\alpha}|^2
+|\mu^\prime_{L\alpha}+\Delta \mu^{\prime}_{L\alpha}|^2) \nonumber \\
&& = M_D (m_2+2\sqrt{m_2 m}+m_3 + m).
\label{sumrulegen}
\end{eqnarray}
As we shall see, the order of the smallness of $m$ is determined from the relic density
and stability of DM, regardless of the DM neutrino mass $M_D$. It is also interesting to
note that the $\theta$-mixing does not depend directly on the neutrino mass constraints,
$m_2$ and $m_3$, so that the dark neutrinos can be produced with tiny mixings to create
the right amount of DM relic density via freeze-in.

Given the active-dark neutrino mixings described above and the aim of establishing them
conventionally, we now propose to perturb the type-I seesaw Lagrangian of
Eq.~(\ref{neutrinos}) by taking in the effective dark couplings
\begin{equation}
- \mathcal{L} \supset \Delta{\it y}^{\prime}_{\nu} \overline{\ell_{L}} \,
\widetilde{\phi} N_{1R} + \Delta{\it y}^{\prime}_L \overline{\ell_{L}} \,
\widetilde{\phi} N^c_{2L} + h.c. ,
\label{darkL}
\end{equation}
maintaining here the interaction with the Higgs field of the SM in order to relate,
after electroweak symmetry breaking, the mass terms of Eq.~(\ref{dev}) to the
corresponding tiny Yukawa couplings, namely,
\begin{equation}
\Delta m^{\prime}_D = \frac{\Delta\it{y}^{\prime}_\nu v_{\phi}}{\sqrt{2}} , \quad
\Delta \mu^{\prime}_L = \frac{\Delta\it{y}^{\prime}_L v_{\phi}}{\sqrt{2}} ,
\label{Deltamass}
\end{equation}
which are analogous to those in Eq.~(\ref{origmass}). However, these $m$ dependent
couplings ($\Delta m_D^\prime = \Delta \mu_L^\prime \sim \sqrt{mM_D}$), darkened by the
seesaw Lagrangian, are not relevant for DM production, dominated by decays of SM weak-gauge
bosons via active-dark neutrino mixing with a relic density independent of the mass
$M_D$ \cite{segregative}. Besides, their contributions to active neutrino masses are
negligible compared to those from the seesaw Lagrangian (see Eq.~(\ref{sumrulegen})),
which conversely, has couplings too large to create the right amount of DM relic
density via freeze-in, as in Eq.~({\ref{TBMmixings}) where there is no place for the
active-dark mixing parameter $m$ ($|m_D^\prime| = |\mu_L^\prime| \sim \sqrt{m_{2,3}M_D}$).

The diagram in Fig.~\ref{fig:Seesaw} illustrates some characteristics of the model:
the Yukawa terms $(\overline{\nu_L} \phi^0) N_{1,2} + h.c.$, obtained from Eq.~(\ref{neutrinos}),
bring in the dark neutrinos; it involves a low-scale dark seesaw with the dark neutrinos
themselves, through which the active neutrinos acquire masses; the dark $Z_2$ symmetry
is not conserved in this dark scenario with vanishingly small abundance of RH neutrinos.
By contrast, there is $Z_2$ conservation in the RH neutrino realm at very high energies,
where the abundance of dark neutrinos is assumed to be null or negligible.

\begin{figure}[ht]
\centering
\includegraphics[width=0.35\textwidth]{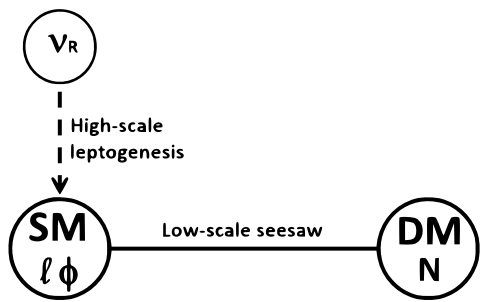}
\caption{Diagrammatic representation of the model: a high-scale leptogenesis with
RH neutrinos and a low-scale seesaw with dark neutrinos.}
\label{fig:Seesaw}
\end{figure}

With the extra sterile neutrinos providing DM and the RH neutrinos decoupled
from low-scale physics, the model can accommodate current experimental and
observational constraints. As a matter of fact, as shown below, interactions
between SM particles and dark neutrinos can reconcile a cold neutrino DM, despite
one of the active neutrinos is massless at tree level. Moreover, the production
and decay of dark neutrinos allow to have the observed relic density with an
upper bound on the DM neutrino mass around 1~MeV.

\section{Phenomenological considerations of the DM neutrino}
\label{Phenomenology}

Since the dark neutrinos are blocked at high energies by RH neutrinos and $Z_2$
symmetry, we assume DM production through the non-thermal freeze-in mechanism. In this
scenario, the interaction of the DM particle with the SM thermal bath is so feeble
that it never reaches thermal equilibrium. The initial DM abundance is negligible,
increasing gradually over the cosmological history via these very weak interactions.
A striking characteristic of the freeze-in production of DM is that allows low-scale
mass DM candidates.

Considering that the mass scale of $N_{1,2}$ is much higher than the mass splitting,
defined by the tiny mass of active neutrinos (see Eq.~(\ref{eigenvalues})), the two
sterile dark neutrinos become approximately degenerate and can be treated as a Dirac
neutrino ($N$) in the following, which then plays the role of DM in the freeze-in
scenario. Constraints from the production and decay of this DM sterile neutrino fix
the allowed parameter space. The DM is expected to be generated non-thermally from
decay of SM particles. In this way, subsequent to the electroweak symmetry breaking,
the DM Dirac neutrino $N$ can be produced from the following dominant decays
\begin{equation}
W^\pm \rightarrow N \ell^\pm , \quad Z \rightarrow N \nu , \quad
\phi^0 \rightarrow N \nu ,
\end{equation}
subject to $M_D < M_W$ and a soft breaking of the $Z_2$ symmetry.
These decays depend on the very small parameters $\Delta m_D^{\prime}$,
$\Delta \mu_L^{\prime}$ in Eqs.~(\ref{darkmixing})-(\ref{Deltamass}), which
characterize the deviation from unitarity of the active neutrino mixing matrix
and the active-dark neutrino mixing. Thermal one-loop corrections that suppress
these mixings at electroweak temperatures, calculated with the thermal quantum field
theory \cite{thermalQFT}, are not taken into consideration because they are not
perturbative at all.

In order to study the evolution of the abundance of the DM Dirac neutrino $N$, denoted by $Y_N$,
until our era, we have to consider its Boltzmann equation involving the temperature $T$
according to \cite{DM1}
\begin{eqnarray}
\frac{dY_N}{dz} &=& \frac{2 M_P}{1.66 m_\phi^2} \frac{z \sqrt{g_\rho}}{g_s}
(\langle \Gamma_{W^\pm \rightarrow N \ell^\pm} \rangle Y_W^{eq}
\nonumber \\ && \nonumber \\
&+& \langle \Gamma_{Z \rightarrow N \nu} \rangle Y_Z^{eq} +
\langle \Gamma_{\phi^0 \rightarrow N \nu} \rangle Y_\phi^{eq}) ,
\end{eqnarray}
where $M_P$ denotes the Planck mass, $z = m_\phi / T$, and $g_\rho$, $g_s$ denote
effective degrees of freedom associated with the energy density and entropy density,
respectively. Here $\langle \Gamma \rangle$ represent the thermally averaged decay
width. Contributions of annihilation processes producing $N$ are subleading to those
of decays and therefore are not put in. Reverse reactions involving $N$ are not taken
into account either, as the $N$ number density is initially insignificant, and for the
same motive, terms proportional to $Y_N$ are discarded too. On the other hand, the
decay of the heavy RH neutrinos at high energies implies that their abundance is
vanishingly small in the electroweak broken phase where $N$ production is emerging
from the decay of the SM weak-gauge and Higgs bosons. Their participation in DM
production is then effectively reduced to zero. Hence, replacing the abundance
$Y_{N}(z_\infty)$ after the freeze-in temperature, the relic density turns out to be,
using standard results~\cite{DM2},
\begin{equation}
\Omega_N h^2 = 2.755 \times 10^5 \left( \frac{M_D}{\mbox{MeV}} \right) Y_{N}(z_\infty) ,
\label{relic}
\end{equation}
where $h$ is the dimensionless reduced Hubble constant.

The effective neutrino mixing parameter
\begin{equation}
|\theta|^2 = \sum_{\alpha=e,\mu,\tau} (|U_{\alpha N_1}|^2 + |U_{\alpha N_2}|^2 ) ,
\label{theta2}
\end{equation}
defined by the effective active-sterile neutrino mixing angles given in
Eqs.~(\ref{darkmixing})-(\ref{Deltamass}), can be estimated from possible decays of
the dark neutrino, which break the $Z_2$ symmetry. The strongest constraint comes
from the radiative decay
$N \rightarrow \nu\gamma$ \cite{decay1,decay2,decay3,segregative} as
\begin{equation}
|\theta|^2 \leq 2.8 \times 10^{-18} \left( \frac{\mbox{MeV}}{M_D} \right)^5 .
\label{Xbound}
\end{equation}
Given that the dominant decays of N involve directly this effective mixing parameter
and, as set in Eq.~(\ref{theta}), $|\theta|^2 = m / M_D$, the final DM relic density,
being related to $M_D Y_N$, is independent of the mass $M_D$, a crucial result that
cannot be gotten from the seesaw Lagrangian alone. This implies that the parameter
$m$ is just fixed by the relic constraint, $\Omega_N h^2 = 0.12$ obtained by the
Planck experiment \cite{DMrelic}. As shown in Ref.~\cite{segregative}, it leads to
the value $m \sim 10^{-12}$ eV, dominated mainly by the decay of SM weak-gauge bosons,
and from Eq.~(\ref{Xbound}) to an upper bound for the DM neutrino mass,
\begin{equation}
M_D = \frac{m}{|\theta|^2} \lesssim 1 \; \mbox{MeV} .
\label{boundMD}
\end{equation}
In Ref.~\cite{segregative}, large thermal one-loop suppressions in active-dark neutrino
mixings at temperatures around the electroweak scale deduced from thermal quantum field
theory~\cite{thermalQFT} were dismissed. We do not take them into account either as the
pertubative scheme breaks down.

Now it is important enough to mention that in the context of relations in
Eq.~(\ref{TBMmixings}), this limit implies
$|m^\prime_{D\alpha}|=|\mu^\prime_{L\alpha}|\lesssim 0.1$ keV, assuming the normal
or inverted ordering for neutrino masses \cite{PDG}, while from Eq.~(\ref{dev}),
$|\Delta m^{\prime}_{D\alpha}| = |\Delta \mu^{\prime}_{L\alpha}|\lesssim 10^{-3}$ eV.
Regarding Yukawa couplings, Eqs.~(\ref{origmass}) and (\ref{Deltamass}) yield
$|\it{y}_\nu^\prime|=|\it{y}_L^\prime|\lesssim 10^{-9}$ and
$|\Delta\it{y}_\nu^\prime|=|\Delta\it{y}_L^\prime|\lesssim 10^{-15}$ (after freeze-in),
respectively. Also, as expected from Eq.~(\ref{darkmixing}), we have active-dark neutrino
mixings which are several orders of magnitude smaller than those obtained from the
approximations $m_{D \alpha}^{\prime}/M_D$ and $\mu_{L \alpha}^{\prime}/M_D$, which
depend on the neutrino mass in Eq.~(\ref{TBMmixings}).

On the other side, we have the so-called Tremaine-Gunn bound \cite{TremaineGunn} which
fixes a lowest mass about 1~keV for the DM sterile neutrino decaying into the X-ray range.
The region of DM neutrino mass is thus between 1 keV and 1 MeV, which restricts the
active-sterile neutrino mixing angle $|\theta|^2$ to be in the extremely small $10^{-15}-10^{-18}$
range. Given these scales, corrections to unitarity of the active neutrino mixing matrix
and to their masses, as in Eq.~(\ref{sumrulegen}), are not very significant. As a matter
of fact, using Eq.~(\ref{findarkmix}), it is predicted that such a mixing matrix remains
unitary at least up to the $\mathcal{O}(10^{-7})$ level \cite{unitarybound}. It is then
worth remarking that in the end the DM relic favored parameter space is consistent with
a feeble active-dark neutrino mixing and the active neutrino mass and mixing.

It is also seen that the lifetime of $N$ (defined by the dominant decay $N \rightarrow 3\nu$)
is longer than the age of the universe and that the non-thermality condition $\Gamma/H < 1$
for the decay channels producing DM is satisfied, where $\Gamma$ is the relevant decay width
measuring the rate of production and $H$ is the Hubble parameter measuring the expansion rate
of the universe at around the temperature $T \sim M$, being $M$ the mass of the decaying particle.
Moreover, the production of $N$ has associated a short free streaming length \cite{DM1,DM3},
qualifying therefore as cold dark matter (CDM) rather than warm dark matter (WDM).

Our proposal has to be contrasted with the more economic $\nu \mbox{MSM}$ model \cite{segregative}
based on a high-scale seesaw mechanism and a severe segregation among RH neutrinos, without
including the extra sterile neutrinos and the $Z_2$ symmetry; similar constraints but on different
setups, both in particle content as in symmetry of the arrangement. In the $\nu \mbox{MSM}$ model
the dark neutrino is of Majorana type with no significant participation in neutrino mass creation,
whereas in our model the dark neutrinos are responsible for the generation of active neutrino
masses and a DM neutrino of quasi-Dirac type. The $Z_2$ symmetry, the decoupling of RH neutrinos,
and the feeble active-dark neutrino mixing in our dark neutrino model assure the stability
of the DM neutrino over the cosmological time scale, which is also guaranteed in the
$\nu \mbox{MSM}$ scenario.

The DM neutrino, being of quasi-Dirac type, is practically irrelevant in lepton number
violating reactions like neutrinoless double beta decays, which would establish the
Majorana nature of active neutrinos. Yet, they can mediate lepton flavor violating
processes such as $\mu \rightarrow e \gamma$, though the active-sterile neutrino mixing
angles are not large enough to be at the range of the present experimental
sensitivity \cite{MEG}. The model also avoids the strong constraints from direct
search \cite{xenon}, as there is no tree-level DM-nucleon coupling. However, dark
neutrinos can be explored at the collider facilities via two-body and three-body
decays \cite{Collider1,Collider2,Collider3,Collider4}.

\section{Conclusion}
\label{Conclusion}

We have proposed an extension of the SM featuring three RH neutrinos, responsible for
baryogenesis via unflavored leptogenesis, and two extra sterile neutrinos, odd under a
$Z_2$ symmetry, in control of the active neutrino masses and providing as well a DM
candidate in the form of a quasi-Dirac neutrino, which behaves as freeze-in type of DM.
The SM regulates the production of dark sterile neutrinos via active-dark neutrino
mixing and the active neutrino masses are generated using a low-scale dark linear seesaw
mechanism. The lightest neutrino is massless at tree level, fixing the absolute neutrino
mass scale to be probed by current experiments \cite{KATRIN,P8}. The non-existence of
dark Majorana mass terms is explained by means of presymmetry instead of supersymmetry as
invoked in the standard linear seesaw, so avoiding to depend on a plethora of undetected
particles and interactions. It is also an alternative to non-supersymmetric versions of
the regular linear seesaw, which require a more involved particle content.

The three RH neutrinos are superheavy due to their huge Majorana masses and their abundance
at the electroweak broken stage vanishingly small because of their decays at high energies,
meaning that they have no impact on the low-scale phenomenology. While their contribution
to active neutrino mass generation via type-I seesaw is strongly suppressed, their decays
work for an unflavored leptogenesis.

The RH neutrinos have no intervention in the making of DM, dominated mainly by the SM weak-gauge
bosons. Current constraints favor a CDM neutrino with mass having an upper bound around 1~MeV,
complying in particular with the relic abundance. The crucial weakness of the active-dark
neutrino mixing is related to the soft breaking of the $Z_2$ symmetry as well as to the tiny breach
of the unitarity of the active neutrino mixing matrix at or below the $\mathcal{O}(10^{-8})$
level, and to an effective minuscule mass value with an upper bound of
$\mathcal{O}(10^{-12})$ eV, perturbing the neutrino masses. Large thermal one-loop corrections
that suppress active-dark neutrino mixings at the time of dark neutrino production, computed with
the thermal quantum field theory, are not taken into account as they are not perturbative at all.
Consistency between the DM favored parameter space and active neutrino mass and mixing is
found. Besides, the feeble coupling of the DM sterile neutrino with active neutrinos is in
consonance with the freeze-in mechanism.

The model was contrasted with versions of the more economical $\nu$MSM model based on a severe
segregation among the three RH Majorana neutrinos, where one of them is accommodated as a light
freeze-in type of DM, whereas the two other superheavy ones are in control of the light neutrino
masses via a high-scale seesaw mechanism and the matter-antimatter asymmetry via flavored
leptogenesis. It was compared with other models that call for the RH neutrinos and relate
neutrino mass generation to DM, and also with extensions of the SM that propose DM sterile
neutrinos below the MeV scale, but with no involvement in the generation of active neutrino
masses. Our approach to explain the mass and mixing of neutrinos, and the genesis of the DM
and baryon asymmetry, is different from all of them, from the point of view of the model and
the phenomenology. The dark linear seesaw itself is minimal given that only two dark sterile
neutrinos are considered, which are responsible for the active neutrino masses and at the same
time provide a DM candidate.

\begin{acknowledgments}

This work was partially supported by Vicerrector\'{\i}a de Investigaci\'on,
Innovaci\'on y Creaci\'on, Universidad de Santiago de Chile (USACH).

\end{acknowledgments}


\end{document}